\documentclass[unnumsec,webpdf,contemporary,large]{oup-authoring-template}%
\graphicspath{{Fig/}}

\theoremstyle{thmstyleone}%
\theoremstyle{thmstyletwo}%
\theoremstyle{thmstylethree}%

\usepackage{makecell}
\usepackage{booktabs}

\begin{document}

\journaltitle{Journal Title Here}
\DOI{DOI HERE}
\copyrightyear{}
\pubyear{}
\access{Advance Access Publication Date: Day Month Year}
\appnotes{Paper}
\firstpage{1}

%\subtitle{Subject Section}

\title[Short Article Title]{Modeling Adaptive Visual Search in Semantically Hierarchical Layouts}

\author[1,$\ast$]{Saku Sourulahti}
\author[1]{Jussi P. P. Jokinen}

\authormark{Sourulahti and Jokinen}

\address[1]{\orgdiv{Faculty of Information Technology, Cognitive Science}, 
\orgname{University of Jyväskylä}, 
\orgaddress{\street{Mattilanniemi 2}, \postcode{40100 Jyväskylä}, \country{Finland}}}

\corresp[$\ast$]{Corresponding author. \href{mailto:saku.p.sourulahti@jyu.fi}{saku.p.sourulahti@jyu.fi}}

%\author[1,$\ast$]{Anonymous}
%\author[2]{Second Author}
%\author[3]{Third Author}
%\author[3]{Fourth Author}
%\author[4]{Fifth Author\ORCID{0000-0000-0000-0000}}

\authormark{Author Name et al.}

%\address[1]{\orgdiv{Department}, \orgname{Organization}, \orgaddress{\street{Street}, \postcode{Postcode}, \state{State}, \country{Country}}}
%\address[2]{\orgdiv{Department}, \orgname{Organization}, \orgaddress{\street{Street}, \postcode{Postcode}, \state{State}, \country{Country}}}
%\address[3]{\orgdiv{Department}, \orgname{Organization}, \orgaddress{\street{Street}, \postcode{Postcode}, \state{State}, \country{Country}}}
%\address[4]{\orgdiv{Department}, \orgname{Organization}, \orgaddress{\street{Street}, \postcode{Postcode}, \state{State}, \country{Country}}}

%\corresp[$\ast$]{Corresponding author. \href{email:email-id.com}{email-id.com}}

%\received{Date}{0}{Year}
%\revised{Date}{0}{Year}
%\accepted{Date}{0}{Year}

%\editor{Associate Editor: Name}

%\abstract{
%\textbf{Motivation:} .\\
%\textbf{Results:} .\\
%\textbf{Availability:} .\\
%\textbf{Contact:} \href{name@email.com}{name@email.com}\\
%\textbf{Supplementary information:} Supplementary data are available at \textit{Journal Name}
%online.}

\abstract{This paper introduces a computational cognitive model to investigate how information grouping impacts visual search, a key consideration in user interface design.
The model uses computational rationality to view user behavior as an adaptation to cognitive and task constraints.
Our work highlights that humans use hierarchical task representations, exploiting semantic and visual structures to improve search efficiency within the constraints of the visual system. 
We validate this model with data from two human studies focused on visual search and semantic categorization, demonstrating that semantic grouping improves search performance when it aligns with spatial grouping.
Our model replicates task durations and eye movement patterns.
By improving understanding of how hierarchical memory structures are utilized in human cognition, the model extends previous visual search models.  
We showcase our model in the rapid prototyping and evaluation of semantic visual groupings within user interface wireframes, suggesting a pathway toward applications in more complex, real-world interface design.
}

% \boxedtext{
% \begin{itemize}
% \item Key boxed text here.
% \item Key boxed text here.
% \item Key boxed text here.
% \end{itemize}}

\maketitle

\section{Introduction}
The organization of information in a user interface (UI) impacts visual search.
Efficient semantic categorizations are known to reduce search times and improve user performance \citep{hornof2001visual, halverson2011computational, salmeron2005expert, chen2015emergence, bailly2014model, ahlstrom2010s, cockburn2007predictive, halverson2008effects, card1982user, mcdonald1983searching, halgren1993towards, brumby2015visual}.
This advantage is particularly pronounced when visual design aligns with interaction semantics, such as by grouping semantically related elements together or positioning related groups in close spatial proximity \citep{brumby2015visual, niemela2003layout, salmeron2005expert}.
Although numerous predictive models of visual search address the impact of structured layouts \citep{halverson2011computational, teo2012cogtool, chen2015emergence, bailly2014model, ahlstrom2010s, cockburn2007predictive}, none yet predict how search behavior adapts to semantic visual hierarchies.

The primary hypothesis for why semantic categorization aids visual search posits that the human visual system exploits global semantic structures \citep{hornof2001visual, halverson2011computational, halverson2008effects, card1982user, mcdonald1983searching, halgren1993towards, brumby2015visual}.
Users can disregard entire groups of elements during a search by encoding one item and determining its mismatch with the target category, provided the group is visually distinct and the search task maintains semantic consistency.
These skills are acquired through usage: the full advantage of semantic categorization becomes apparent only with prolonged exposure to consistently organized groups, making structured visual search a highly adaptive process \citep{mcdonald1983searching, brumby2015visual}.
Yet, the question remains: how does such adaptation arise from specific spatial arrangements of semantically related items?

\begin{figure*}[htb]
    \centering \includegraphics[width=1.0\textwidth]{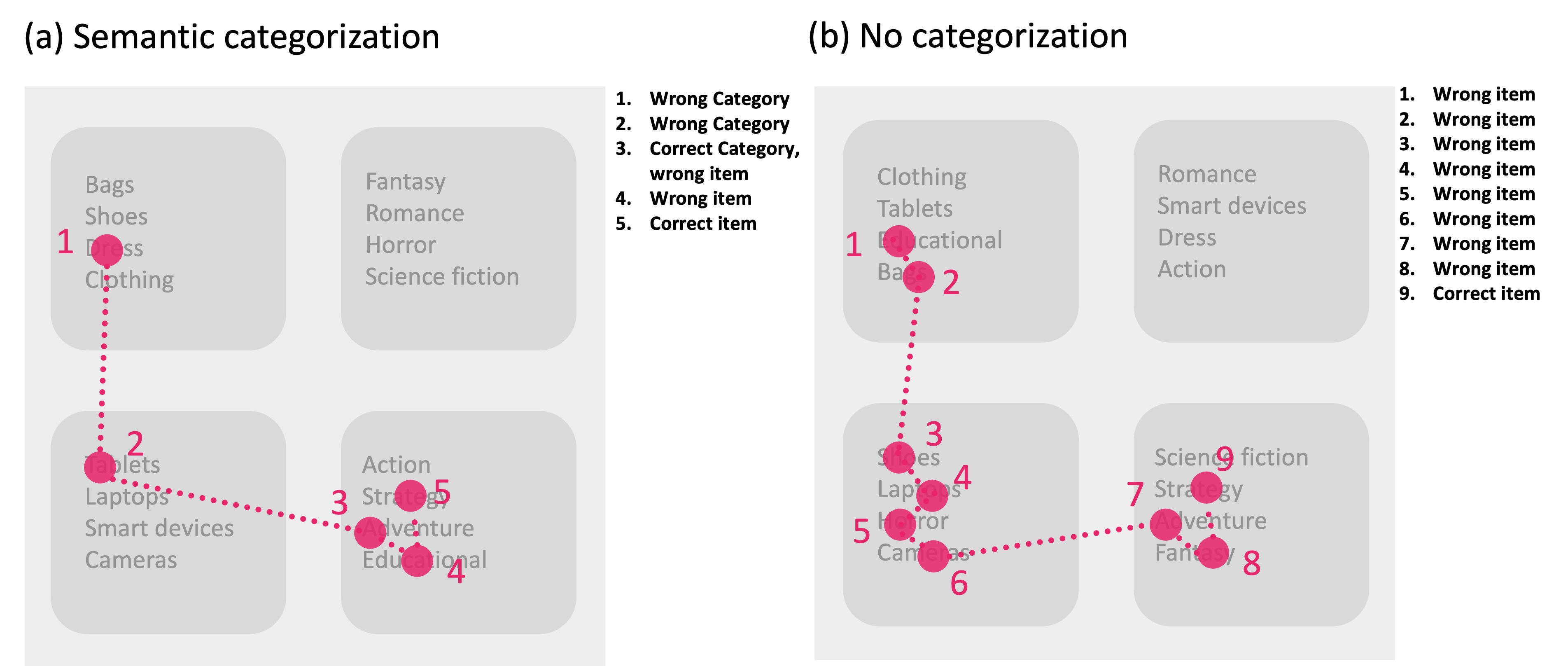}
    \caption{Our model simulates step-by-step fixations during visual search. (a) When item groups are semantically organized, our model jumps over groups that semantically mismatch with the search target. (b) Without semantic consistency in visual groups, our model is forced to exhaustively search through visual groups.
}
\label{fig:eye_path}
\end{figure*}

%This paper explores the concept of adaptation through \emph{computational rationality}, which explains the emergence of visual search strategies in light of the cognitive capacity of the user and the constraints of the search task.
Figure \ref{fig:eye_path} illustrates this problem and demonstrates how our model approaches it.
In pane A, each visually distinct category includes elements belonging to a common semantic category.
Our model anticipates this arrangement, enabling it to bypass entire visual areas by assessing the semantic distance between the search target and a single observed item \citep{brumby2015visual}.
Upon encountering a visual group with a matching semantic category, the model initiates a detailed, element-by-element search, thereby optimizing visual search efficiency.
A different search behavior arises with designs lacking clear semantic categorization (pane B, Figure \ref{fig:eye_path}).
%The model cannot bypass entire groups of elements due to inconsistent semantics, resulting in longer search times.
The model cannot make use of consistent semantics, but it can still utilize visual structures in the task.
By methodically searching all elements within one visual group before moving to the next, the model addresses a limitation in human visual short-term memory: the capacity to remember only a small number of previously visited elements to avoid revisiting them \citep{sourulahti2024visual}.
This emphasizes the adaptive nature of our model: it develops a search policy that optimally uses the available cognitive resources, given the visual design of the search task.

Models based on \emph{computational rationality} have shown that human-like behavior emerges by framing it as bounded goal-directed adaptation, and by computationally solving the problem of how to optimize behavior within cognitive and environmental constraints \citep{chen2015emergence, jokinen2020adaptive, jokinen2021multitasking, jokinen2021touchscreen, oulasvirta2022computational}.
In this work, we adopt this approach to model how human visual search adapts to the presence of semantic and visual hierarchical structures.
The advantage of this approach lies in its ability to reveal unexpected search strategies, which makes the model explanatory rather than merely predictive \citep{oulasvirta2022computational}.
By doing so we consider a key gap: earlier models do not predict how search strategies develop based on cognitive capacity limitations and semantic task environment structure, which is the main objective of our research.

To date, visual search models have demonstrated limited ability to generalize to semantically and visually complex UI environments.
Progress is hindered by the need to explicitly model each specific behavioral phenomenon arising from cognition, whether via production rules \citep{teo2012cogtool, halverson2011computational} or through mathematical formulations \citep{cockburn2007predictive}.
Rationally adaptive modeling offers a more robust approach to search-model development by grounding human behavior with more fundamental cognitive principles. 
This moves the field closer to the objective of simulating visual search in real-world user-interface environments.

Our modeling approach demonstrates that visual search strategies emerge from low-level visual features and from implicit hierarchical information within the environment: the cognition exploits these to efficiently support task-related objectives. 
The work presented in this paper contributes to the theoretical understanding of the cognitive processes underlying visual search, enabling more accurate predictions of gaze paths in hierarchically complex visual environments, such as UIs. 
Furthermore, the model offers practical value by supporting designers in evaluating how visual structures and semantic relationships jointly facilitate efficient information retrieval.
We make the following contributions:
\begin{itemize}
\item A formalism of a computational cognitive model of visual search that predicts how search adapts to the presence of visual and semantic structures.
\item Validation of the model with two experiments, one using original data and another using data from a previous experiment.
\item We provide the model as open access, permitting researchers to further develop visual search models and designers to explore how their visual design wireframe choices interact with human cognition.
\end{itemize}

%% The goal of this study is to: 

%% 1.	Determine, using a predictive model and empirical results, whether search strategy improves in semantically hierarchical layouts as a result of adaptability.

%% 2.	Determine, using a predictive model and empirical results, how the constraints of semantic categorization together with spatial structures, affect search performance.

%% 3.	Develop a tentative tool to assist HCI design in evaluating how semantically grouped layouts impact search performance.

\section{Related work}\label{sec2}
Computational models of semantically hierarchical visual search are typically \citep{jokinen2020adaptive} categorized into models that simulate eye movements \citep{teo2012cogtool, halverson2011computational}, and mathematical models that predict search outcomes \citep{cockburn2007predictive, bailly2014model}.
These models often employ a fixed set of search strategies, reliant on predefined logical conditions to guide gaze deployment.
The Search-Decision-Pointing (SDP) models \citep{cockburn2007predictive, bailly2014model} estimate search performance based on external factors such as menu length, target position, and the user's knowledge.
Adjusting these parameters aids designers in understanding design impact on search; however, they overlook the cognitive processes integral to search. %, like the mechanisms driving eye movements.
Consequently, although they efficiently predict search times, these models are largely empirical, estimating search durations based on physical layout properties rather than providing causal cognitive explanations \citep{murray2022simulation}.

Some models explore how a hierarchical approach improves search performance in spatially and semantically grouped visual areas \citep{teo2012cogtool, halverson2011computational}.
These are developed within cognitive architectures, such as ACT-R \citep{teo2012cogtool} and EPIC \citep{halverson2011computational}.
Their strength lies in being grounded on psychological hypotheses regarding human cognitive processes, abilities, and limitations.
This includes detailed accounts of eye movements, such as saccade preparation and execution times \citep{salvucci2001integrated}, capacities like limited visual short-term memory (VSTM) for inhibiting return to previously visited elements \citep{posner1985inhibition}, and the deployment of attention to visually salient elements \citep{jokinen2020adaptive}.
However, these models often rely on predefined rules governing eye movement strategies, which may oversimplify the decision-making involved in global search \citep{oulasvirta2022computational}.
This reliance limits their ability to explain how strategies emerge from cognitive constraints, such as VSTM, and task-environment dynamics, such as spatial and semantic groupings.
To date, existing models have not demonstrated how the proposed search strategy for spatial and semantic categorization \citep{brumby2015visual} can arise as an adaptation to these constraints and dynamics.

Computational rationality offers a framework for understanding behavior as an optimal adaptation to the environmental limits and cognitive capacities \citep{gershman2015computational, lewis2014computational, oulasvirta2022computational}, and models based on it have recently been a popular approach for modeling interaction \citep{acharya2018automation,cheema2020predicting,chen2017cognitive,jokinen2021multitasking,jokinen2021touchscreen,lingler2024supporting}.
An \emph{agent} is tasked with maximizing the long-term gain of sequential actions within a given environment.
This agent does not follow a predefined set of rules but learns to associate environmental and cognitive states with appropriate actions.
Via reinforcement learning \citep{sutton2018reinforcement}, the agent adaptively learns to optimize behavioral performance by mapping actions to its cognitive representation in a way that maximizes long-term time-discounted cumulative return.
Models based on production rules provide transparency in how decision making occurs, but necessitate the modeler to explicitly hypothesize the logic of this process.
Conversely, computational rationality gives an explanatory account of how decisions emerge under cognitive constraints.
In other words, this hypothesis of optimal adaptation is foundational, and the logic of decision making is postulated to emerge in a principled way from the foundation.

The model presented in this paper builds upon a recent visual search model that proposes representing search environments hierarchically to overcome the capacity limitations of VSTM \citep{sourulahti2024visual}.
It is grounded in the multi-level representational theory of visual memory \citep{brady2011hierarchical, brady2013probabilistic, graham2000traveling, kong2010high}.
The role of capacity in visual memory becomes particularly pronounced in large-scale search tasks, especially when visual search is not guided by bottom-up salience or top-down control \citep{sourulahti2024visual}. 
In such task environments, the search strategy is characterized by an attempt to locate the target efficiently in a new location, a phenomenon known as inhibition of return \citep{posner1984components, posner1985inhibition}. 
The empirically observed robust visual processing of large numbers of items, however, cannot be explained by previously proposed models of visual working memory, whether based on a limited number of encoded items or features \citep{cowan2001magical, luck1997capacity, vogel2001storage}, or on shared-resource memory \citep{bays2009precision, franconeri2013flexible, 
gorgoraptis2011dynamic,  ma2014changing}. 
VSTM must also be able to organize visual information into chunks, allowing it to encode the locations of entire regions rather than merely single elements \citep{miller1956magical, woodman2003perceptual}
The structure of the visual memory must therefore be flexible, allowing it to be organized in a hierarchical manner \citep{brady2011hierarchical, brady2015contextual, brady2013probabilistic, kong2010high}, in order to adapt optimally to the visual structures of the task environment \citep{chen2015emergence, jokinen2021multitasking, jokinen2020adaptive, todi2019individualising}.
To make use of this, search policy must aim to exploit the visual and semantic grouping of the environment for efficient gaze deployment, operating both globally across groups and locally within groups 
\citep{brady2011hierarchical, brady2015contextual, brady2013probabilistic, kong2010high, nakashima2013visual, xu2010impact}

Where previous models have extended the VSTM buffer to hold more elements in order to make it possible to simulate complex visual search \citep{jokinen2020adaptive}, our model implements a second hierarchical level.
Encoded elements are added into a first-in-first-out buffer with a fixed length, and the agent learns to inhibit fixations to these elements.
At the lowest level of the hierarchical representation, inhibition applies to individual visual elements.
At a higher level, visually identifiable structures of such items facilitate the creation of groups.
Once all elements within such a group have been examined, the entire group can be internally flagged as searched.
This means that, instead of retaining all its individual elements in VSTM, the group itself is treated as a visual ``superelement'' and inhibited.

Building on this concept of hierarchical VSTM, our work utilizes a previously established theory of hierarchical search \citep{sourulahti2024visual} to include semantic groupings. 
The hierarchical search strategy of the model presented in this paper emerges as a result of adaptation, when the visual system learns to exploit the semantic organization of the environment \citep{brumby2015visual}. 
The model adapts to utilize environmental information both when semantic grouping aligns with visual grouping and when it does not.
This means that the hierarchical VSTM contained in the model representation enables the encoding of relevant and irrelevant search areas for target search.
Although bottom-up salience and top-down cues were minimized in our study’s search-task environment, recognition of semantic category membership can still be expected to compete with an item’s orthographic properties in guiding attention \citep{leger2012orthographic}. 
Thus, although category membership can serve as a cue to guide target search, the mechanism is noisy; when semantic information is unavailable, adaptation relies on a structured search strategy.

The model is also limited by external environmental factors.
The structured VSTM allows the model to to reduce the space of possible search policies, but this process is constrained by the degree to which categorization aligns with spatial grouping.
In this study, we investigate how visual-semantic alignment affects visual search performance and search strategy: when users anticipate
that a particular visual group contains elements sharing a semantic group, inhibiting that visual group only necessitates
verifying that any of its elements belong to a different semantic group than the search target \citep{brumby2015visual,halgren1993towards,hornof2001visual,halverson2011computational, halverson2008effects,card1982user,mcdonald1983searching}.

\section{Model description}\label{sec3}
\subsection{Overview}\label{subsec1}

\begin{figure*}[htb]
    \centering
    \includegraphics[width=1.0\textwidth]{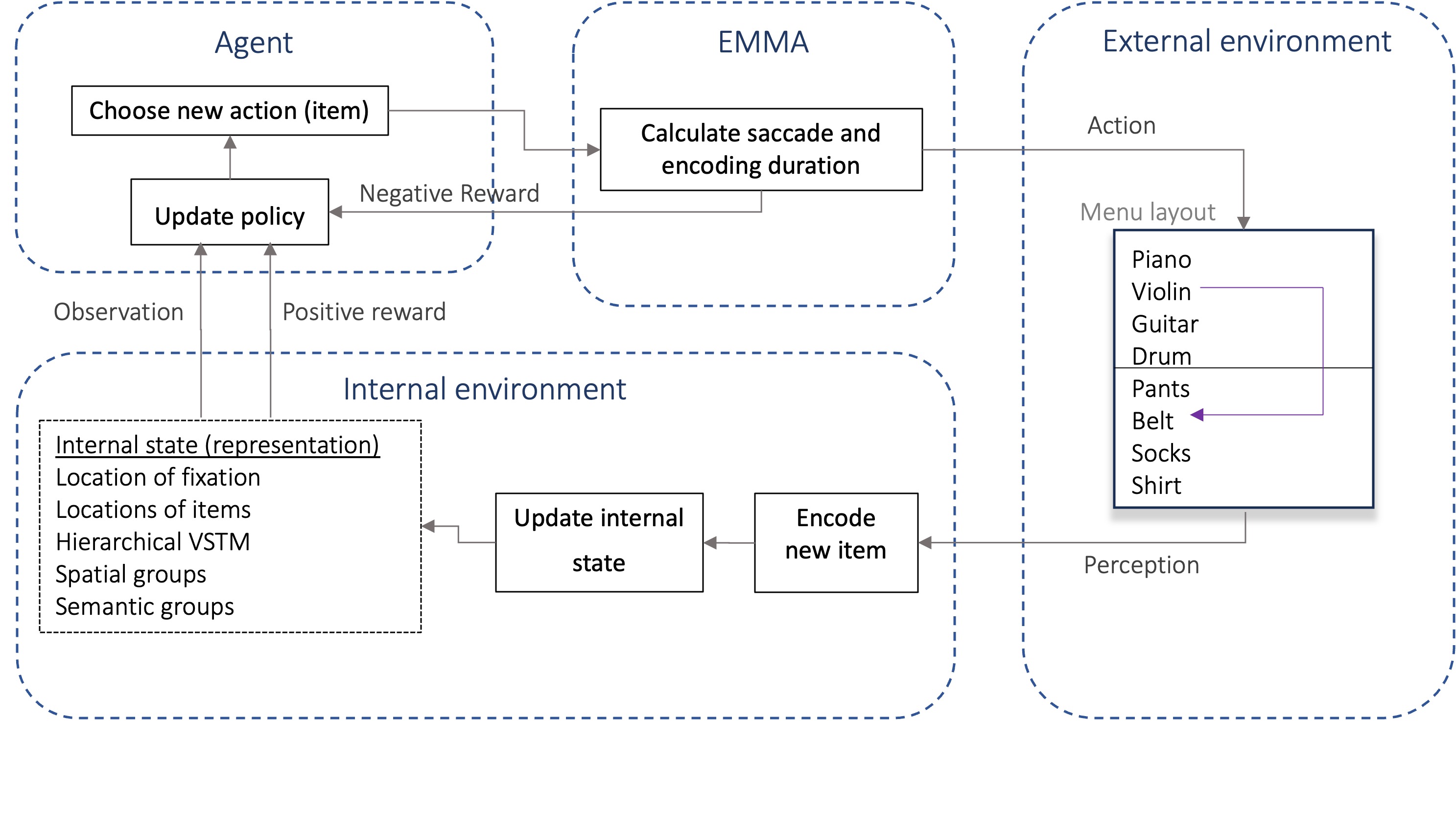}
    \caption{Overview of the interaction between the model's internal and external environments: The agent has adopted a policy that selects eye movement actions maximizing long-term cumulative rewards.
      An action to encode a element is evaluated by the EMMA model, which returns a negative reward based on encoding time.
      A new fixation is deployed in the external environment, which generates an observation.
      The observation updates the representation in the internal environment, which the agent utilizes to decide on the next eye movement action.
      Upon successful encoding of the search target, a positive reward is signaled to the agent.}
    \label{fig:fig_arch}
\end{figure*}

Computational rationality analyzes interaction into an external task environment, internal cognitive environment, and an agent \citep{chandramouli2024workflow,oulasvirta2022computational}.
Figure \ref{fig:fig_arch} depicts our model: the external environment consists of a UI, or a number of rectangular visual elements that have position, size, and text content.
The agent's goal is to deploy eye movements to visually encode the search target.
Given the agent's decision of where to fixate next, an eye movement model computes movement time and simulates a refixation.
The fixated element is encoded into an internal representation, which the agent observes and decides on the next action.
The search policy of the agent is adapted to make use of hierarchical representations in the internal environment, finding search targets even when the number of visual elements greatly exceeds the capacity of VSTM.

%At the heart of  decision-making is optimally controlled, which means that the agent adjusts a sequence of successive actions to be optimal for achieving the goal. The model's decision-making can be thought to follow the so-called Partially Observable Markov Decision Process (POMDP), in which the model moves through successive steps between states, with each subsequent step depending on the previous one \citep{oulasvirta2022computational}. POMDP is derived from the Markov Decision Process (MDP), but in POMDP, the agent receives only partial observations from the environment, where decision-making depends solely on the belief state of the internal representation rather than directly on the actual state of the environment.

%The goal of the adaptive model's learning is to maximize the reward defined by the task, guiding the agent to optimally control decision-making in a direction that minimizes eye movement search time as well as possible. Decision-making forms from observations of the changing states of the external environment, but implicitly through the representational state of the internal environment. This means that the internal environment contains a psychological representation of the external state with defined cognitive constraints, making the model’s input symbolic. The agent optimizes decision-making for the next fixation based on the received reward. Once the agent has made the decision for the next eye movement fixation, it causes a change in the external state. 

% Place the figure right here with the [H] specifier

\subsection{Optimal Control of Eye Movements}\label{subsec2}

% our model is \emph{agency}: an agent selects subsequent actions based on computed values of potential outcomes.
%It adapts action choices to internal representations and environmental constraints.
%We hypothesize that human visual search is optimally controlled through eye movements: choices to deploy visual attention and encode elements are optimal, given the internal and external environments.
%To test this hypothesis, we use the approach of employing reinforcement learning (RL) to develop computational rational models of human behavior \citep{acharya2018automation,cheema2020predicting,chen2017cognitive,jokinen2021multitasking,jokinen2021touchscreen,lingler2024supporting}.

Our model is formalized as a partially observable Markov decision process (POMDP) \citep{kaelbling1998planning,sutton2018reinforcement}, represented as the tuple $<S, A, T, R, \Omega, O, \gamma>$.
At each time step, the model exists in a state $s \in S$.
The agent performs an action $a \in A$, transitioning to a new state $s' \in S$ according to the transition function $T(s, a, s') = p(s' | s, a)$.
This yields a real-valued reward $r \in \mathbb{R}$, determined by the reward function $R(s') = r$.
Due to perceptual limitations such as foveated vision, the agent does not fully observe $s'$, but instead receives an observation $o \in \Omega$ via the observation function $O(s', a, o) = p(o | s', a)$, where $\Omega$ is the set of all possible observations.
Through RL, the agent learns to maximize the expected long-term discounted reward ($\gamma \in [0,1]$), by following an optimal policy $\pi^*$.
A policy $\pi(s, a) = p(a|s)$ assigns probabilities to actions given states, and an optimal policy maximizes the value function:

\begin{equation}
  \label{eq:bellman}
  V^*(s) = \max_a \left[ R(s, a) + \gamma \sum_{s' \in S} T(s, a, s') V^*(s') \right].
\end{equation}

The recursion in Equation \ref{eq:bellman} compels the agent to explore the environment through trial-and-error, leading to gradual convergence towards an optimal policy.
In computational rational modeling, the learning method of the policy is not necessarily assumed to mimic human learning; however, once an optimal policy is achieved, the adapted behavior should be approximately human-like, provided that the internal cognitive environment and the task environment are accurately modeled \citep{chandramouli2024workflow}.
In this paper, we concentrate on modeling the internal cognitive environment to simulate how humans exploit hierarchical semantic structures in visual search, thereby overcoming limitations in VSTM.

In our model, the state describes the visual elements within the graphical layout and the current fixation point.
The action space allows the agent to fixate on any individual element, initiating a simulated eye movement and subsequent encoding of the element's content, as governed by the transition function.
We employ the EMMA eye movement model to calculate the duration of this saccadic movement \citep{salvucci2001integrated}, assigning its duration as a negative reward to the agent (see Appendix A).
The contents of the encoded elements are stored in a VSTM buffer within the internal environment through an observation function.
Additionally, a hierarchical representation of the search task is updated.
The observation function also evaluates the semantic category of the encoded element in relation to the search target's semantic category.
The model code is available in the anonymized repository \citep{anonymous2025osf}.

\subsection{The Internal Representation}\label{subsec3}

In this paper, we concentrate on how visual search adapts to visual and semantic structures, rather on the low-level details of perceptual processes that construct these visual structures, or the semantic information processing that determines distances between labels.
In other words, the representation in our model is formed symbolically \citep{teo2012cogtool, halverson2011computational, chen2015emergence, navalpakkam2005modeling, pomplun2003area, sun2008computer, sourulahti2024visual}. 

In our model, perceptions of the external environment are provided by the observation function to the internal representation.
Visual groups are defined by their location within a search task, with item placement enforcing visual separation between different groups.
Additionally, semantic distances between labels are based on common knowledge, grouping labels into the same or different semantic categories.
First, the observation function defines visual groups based on a section of the search task which they inhabit, with item placement observation visual separation between items of different groups.
Second, semantic distances between labels are provided from a dataset that either matches or does not match labels to the same semantic category.
In the last section of this paper, we discuss future steps to implement more fine-grained, bottom-up models of how this internal representation could emerge in the human cognition.
Here, we assume that such solutions can in future produce the representations presently in use, and focus on the impact of such representations on visual search adaptation.

The information in the model’s internal representation consists of
1) the location of the current fixation,
2) distances between the current fixation and all the other items,
3) a memory buffer for storing previously visited items and item groups,
4) the spatial groups of items, and
5) the semantic relevance between presently encoded item and the search target.
The agent learns to deploy attention efficiently by using this information.
It exploits the lower VSTM level to inhibit return to recently encoded visual elements \citep{posner1984components, posner1985inhibition}.
However, this buffer is limited in size \citep{cowan2001magical, luck1997capacity, vogel2001storage}, so it alone cannot explain how humans are able to carry out efficient visual search of tasks that may contain tens or even a hundred elements.

To enable search with large set sizes, our model’s ability to learn and utilize hierarchical information is based on a two-level VSTM structure. 
This higher VSTM level marks visually distinctive groups of items when all items of that group have been encoded.
Now only one memory element in the buffer is needed to inhibit return to all visual elements in the group, facilitating memory management in complex tasks.
A global search strategy emerges when the agent learns to skip categorically irrelevant list groups and exhaustively search through only those that match the semantics of the search target.
However, this mechanism is limited and inherently noisy with respect to whether category-consistent items are encoded into memory \citep{leger2012orthographic}.
Importantly, our model does not have heuristic rules for inhibiting return to marked elements or groups, or describing how to make use of the semantic information.
The agent learns to utilize information in the internal representation for optimal control of eye movements.

\begin{figure*}[htb]
    \centering
    \includegraphics[width=1.0\textwidth]{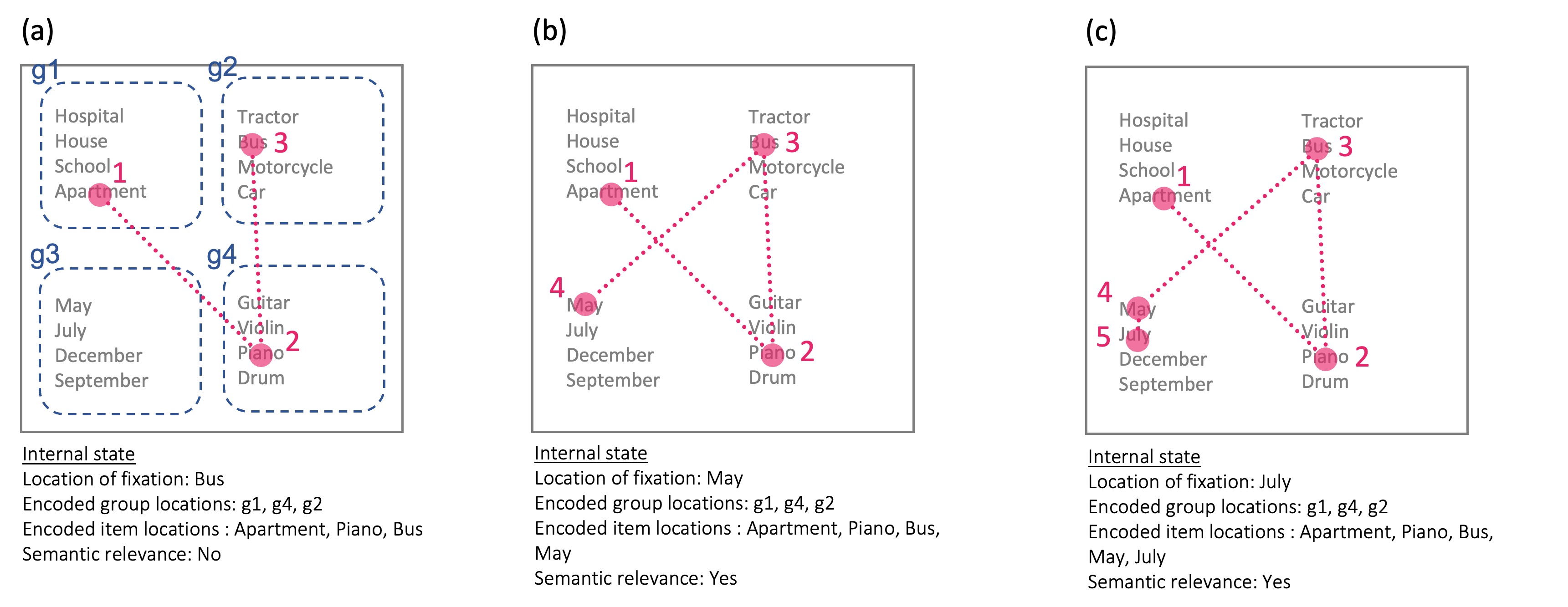}
    \caption{The progression of the model’s internal state, fixation by fixation, demonstrates how it leverages the semantically structured layout. The scan path, direction of fixations, and fixation count are highlighted in red. The model’s internal state is depicted below the layout.}
    \label{fig:fig_rep}
\end{figure*}

Figure \ref{fig:fig_rep} illustrates the progression of our model's internal representation, fixation by fixation, in a task where visual and semantic groupings are aligned.
The illustration excludes task-specific information on element distances and visual structures, focusing on more the dynamic part of the internal representation.  
%(for the full observation space of our agent, see Appendix B).
In this simulation example (Figure \ref{fig:fig_rep}), the target is the label ``July''.
Pane (a) shows areas delineated by blue dashed lines, which indicate elements belonging to visually salient list groups (g1, g2, g3, and g4).
The internal representation includes information on the current fixation location, the locations of groups and items encoded in memory, and the semantic relevance of the fixated item to the target category.
This example is an abstract simplification of the model’s actual computational state change within the internal representation. 
For the full data structure of the observation space and its updating, see Appendix B.
At this initial stage, the search task has advanced through three fixations, with the fixation currently on the item labeled ``Bus.''
The groups g1, g4, and g2 are stored in memory, as their semantic category is not related to the target's semantics.
Since the current list group is also semantically irrelevant, the model advances to the next list group that has not yet been searched.
In pane (b), the search progresses to the fourth fixation, encoding ``May'' into memory.
Although it does not match the search target, ``May'' is semantically in the same category, warranting continued search within this list group.
Pane (c) depicts the fifth fixation, where the model encodes ``July,'' matching the search target and concluding the task.

The model discussed in this paper builds upon an existing model that uses hierarchical representation to predict how search adapts to visual structures \citep{sourulahti2024visual}.
The major improvement in our model is its capability to inhibit visually hierarchical structures by encoding a single item, aided by semantic information.
We retain the existing hierarchical framework and augment it with a semantic layer, allowing the determination of whether the visual group of the encoded item is semantically close to the target item being searched for.

\section{Experiment 1}\label{sec4}
\subsection{Method}

\subsubsection{Design}
The first evaluation of our model focuses on visual search time, which in our case serves as a general metric for how visual design influences search behavior.
To that end, we developed an online visual search task where participants located target items on the screen.
We employed a $3 \times 3$ factorial mixed design for the experiment.
The first design variable, semantic order (between-subjects factor), included three conditions: (1) Unique categorization, with each visual group containing a distinct category; (2) Shared categories, where two visual groups shared the same semantic category; (3) No categorization, lacking semantic consistency among visual groups.
The second design variable was set size (within-subjects factor), with tasks containing either 16, 24, or 48 visual items.
The set size order was randomized for each participant.
In the three between-subjects semantic order groups, participants completed a total of 45 tasks with the three different within-subjects set sizes.

\subsubsection{Participants}
A total of $N=60$ participants were recruited via the Prolific online crowdsourcing platform.
%Requirements for participation included being native English speakers aged between 18 and 29 years, possessing normal or corrected-to-normal vision, and having no physical or cognitive disabilities that could adversely affect visual search performance.
To ensure that participants had no diminished reaction capabilities and they understood the instructions, the requirements for participation included being native English speakers aged 18 to 29. Participants were also required to have normal or corrected-to-normal vision and no physical or cognitive disabilities that could adversely affect visual search performance.
Compensation for participants was in accordance with platform standards, and the experiment conformed to the ethical guidelines of the hosting institution.
All participants provided informed consent prior to participation.

\subsubsection{Stimuli and Procedure}
%The study's stimuli consisted of visual search tasks featuring visually categorized lists of words.
In the study, the stimuli items consisted of single-word labels and were spatially arranged into four-item lists that were either categorized or randomly ordered. The suitability of the word list categories was pre-tested prior to the study and included easily recognizable, distinct categories such as animals, sports, clothing, countries, instruments, and furniture.
Participants were tasked with searching these lists for a specified target word.
Following task instructions, participants completed nine practice tasks before proceeding to the 45 main tasks.
Each task began with a 4-second preparation period during which participants were instructed to position their finger on the reaction button (space bar).
This was followed by a 3-second display of the target cue on the screen.
Subsequently, the stimulus displaying the word lists appeared, marking the start of the task.
Upon locating the cued target, participants pressed the reaction button, which recorded the search time and initiated the next task.

\subsubsection{Data Analysis and Modeling}
A total of 2700 trials were collected from the 60 participants in the study.
Trials with search times larger than 1.5 times the IQR above the third quartile (Q3) or below the first quartile (Q1) were considered outliers.
After outlier removal, a total of 2513 trials (93\%) remained for statistical analysis.

For analyzing within-condition aggregate search times, we used multilevel regression (\texttt{lme4} package in \texttt{R}) with search time as the dependent variable, semantic categorization and set size as independent variables, and their interaction included as a fixed effect. 
Participant was included as a random intercept.
Individual variation accounted for only a small portion of the total variability in the search times, adjusted $ICC = 0.11$.

To fit the model to the results of Experiments 1 and 2, we adjusted the model’s memory capacity to encode categories with a 20\% probability.
In addition, we fine-tuned the EMMA model by adding a 125 ms increment to the preparation time.
Fit between human and model search performance was assessed using $R^2$ and $RMSE$.
Moreover, we tested our model against a baseline strategy that always fixates on the nearest unvisited item.
This baseline is provided with unlimited VSTM capacity for inhibiting return, as otherwise the search task would be infeasible \citep{jokinen2020adaptive}.
The baseline model’s visual environment corresponds to the random condition of the adaptive model. 
This comparison with the baseline model simulations exhibit that humans can perform very close to the theoretical minimum, despite not having unlimited memory \citep{jokinen2020adaptive}. 
The prediction errors of the baseline and adaptive models are evaluated using the Mean Absolute Percentage Error (MAPE).
The experimental data and analysis scripts for Experiments 1 and 2 are available in the anonymized repository \citep{anonymous2025osf}.

\subsection{Results}

Estimated marginal mean search times for each condition, along with 95\% confidence intervals, are shown in Figure 4.
As expected, set size significantly impacted search times: participants took significantly more time to respond as the number of items increased $F(2,2447) = 425.0, p<0.001$.
Likewise in line with our expectation, categorization appears to have a significant effect on search performance $F(2,55) = 12.6, p<0.001$.
Finally, an interaction effect between set size and semantic order condition was observed $F(4,2447) = 12.2, p<0.001$.

\begin{figure}[htbp]
    \centering
    \includegraphics[width=0.5\textwidth]{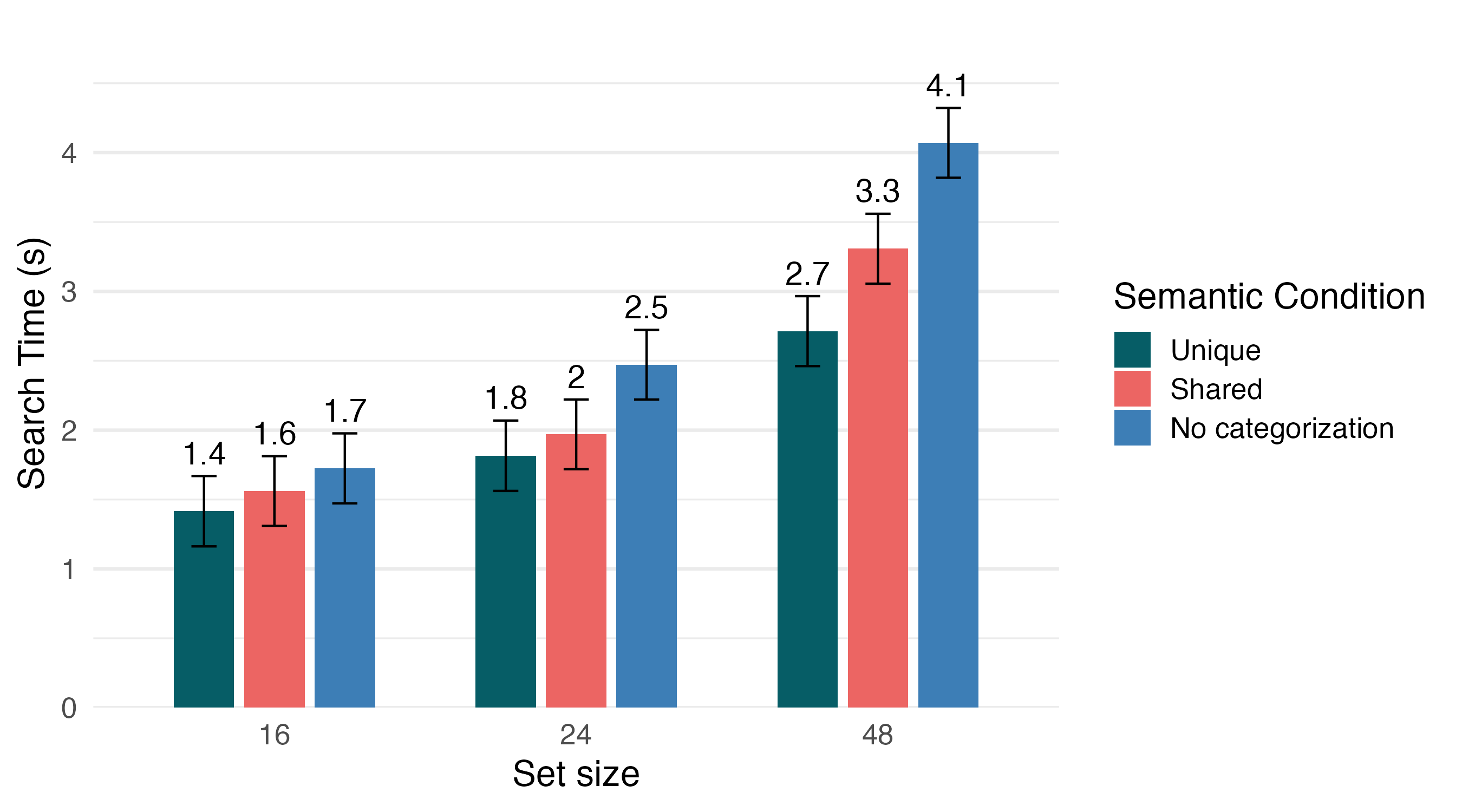}
    \caption{Estimated marginal mean search times across study conditions. Error bars indicate 95\% confidence intervals.}
    \label{fig:fig_emmeans}
\end{figure}

\begin{figure}[H]
    \centering
    \includegraphics[width=0.45\textwidth]{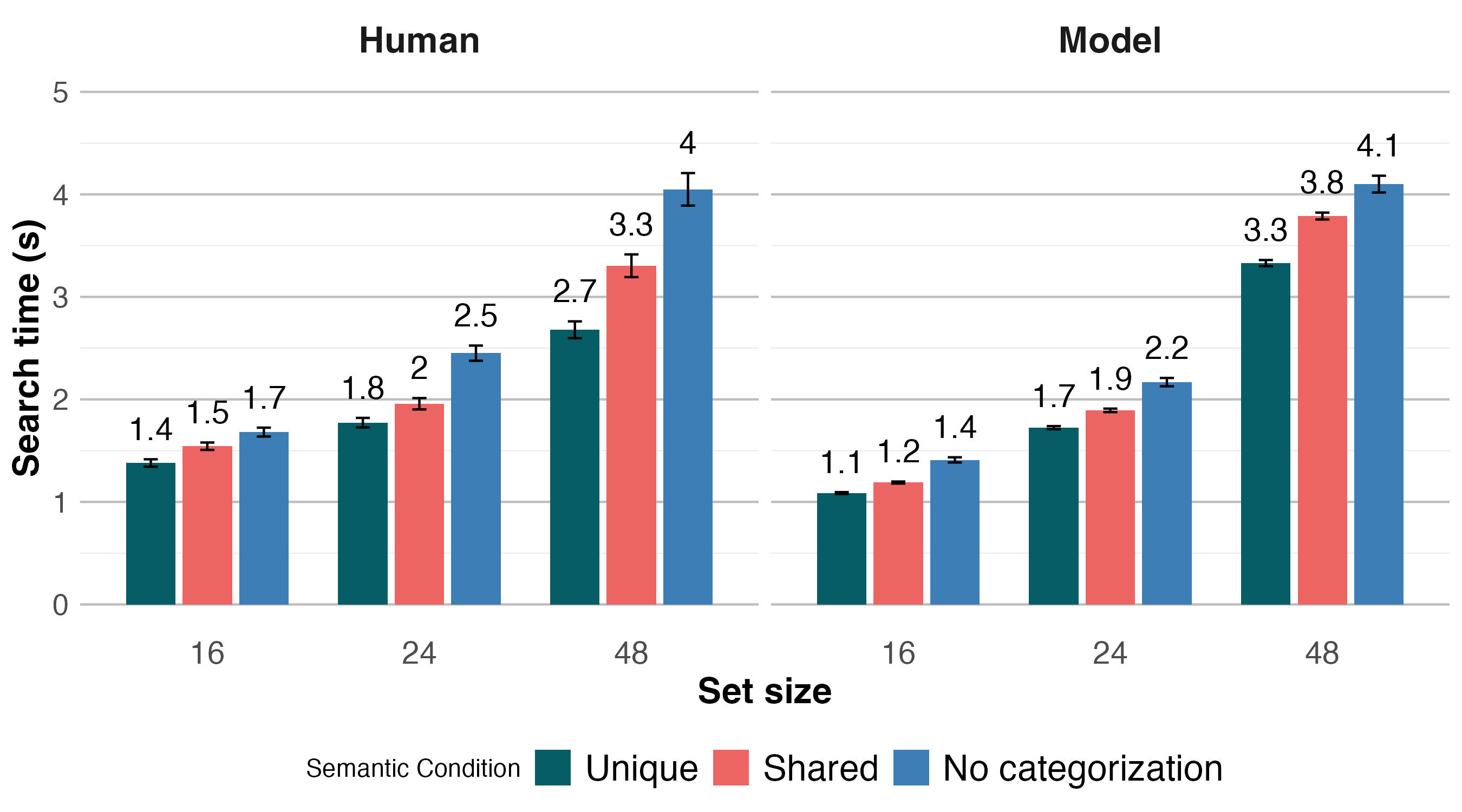}
    \caption{Average search times for adaptive model and humans data between semantic order and set size. The error bars are of similar magnitude across conditions.}
    \label{fig:fig_lines}
\end{figure}

The consistency of the model’s predictions with human search time is shown in Figure \ref{fig:fig_lines}, as well as the means and standard deviations in Table \ref{tab:time_mean_st}. The model's mean absolute percentage error (MAPE) relative to human search time are presented in same Table \ref{tab:time_mean_st}.
Model fit was $R^2 = 0.94$, $RMSE = 0.2 s$.
The model predicted successfully the impact of categorization and set size on search time, as well as their interaction.
The discrepancy is most evident at the largest set size, where the model slightly overestimates search time, and at the smallest set size, where it tends to underestimate it.

\begin{table}[H]
\centering
\resizebox{\columnwidth}{!}{%
\begin{tabular}{l l l l l l l}
\toprule
\makecell[lt]{\textbf{Set}\\\textbf{Size}} &
\makecell[lt]{\textbf{Semantic}\\\textbf{Order}} & 
\makecell[lt]{\textbf{Human}\\\textbf{Mean (s)}} & \makecell[lt]{\textbf{Human}\\\textbf{SD (s)}} & \makecell[lt]{\textbf{Model}\\\textbf{Mean (s)}} & \makecell[lt]{\textbf{Model}\\\textbf{SD (s)}} & \makecell[lt]{\textbf{MAPE}\\\textbf{(\%)}} \\
\midrule
16 & No categorization & 1.7 & 0.7 & 1.4 & 0.8 & 16.1 \\
24 & No categorization & 2.5 & 1.3 & 2.2 & 1.3 & 11.5 \\
48 & No categorization & 4.0 & 2.7 & 4.1 & 2.6 & 1.26 \\
16 & Shared            & 1.5 & 0.6 & 1.2 & 0.7 & 23.0 \\
24 & Shared            & 2.0 & 0.9 & 1.9 & 1.2 & 3.29 \\
48 & Shared            & 3.3 & 1.9 & 3.8 & 2.4 & 14.7 \\
16 & Unique            & 1.4 & 0.6 & 1.1 & 0.7 & 21.3 \\
24 & Unique            & 1.8 & 0.8 & 1.7 & 1.1 & 2.76 \\
48 & Unique            & 2.7 & 1.4 & 3.3 & 2.1 & 24.3 \\
\bottomrule
\end{tabular}
}
\caption{Mean, standard deviation (SD), and mean absolute percentage error (MAPE) of search times for human and adaptive model data across all semantic orders and set sizes.}
\label{tab:time_mean_st}
\end{table}

The baseline model with unlimited memory outperforms the adaptive model under randomized conditions across different set sizes, as expected.
In the baseline model, RMSE was $0.37\,\mathrm{s}$, and at the largest set size, the MAPE exhibited a higher error rate (Table \ref{tab:tab_base})) compared to the adaptive model results (Table \ref{tab:time_mean_st}).
Its fixation count is slightly lower (Table \ref{tab:tab_base}) than that of the adaptive model in the random condition (Table \ref{tab:mod_adap}).

\begin{table}[H]
\centering
\resizebox{\columnwidth}{!}{%
\begin{tabular}{l l l l l l}
\toprule
\makecell[lt]{\textbf{Set}\\\textbf{Size}} & 
\makecell[lt]{\textbf{Mean Search}\\\textbf{Time (s)}} & 
\makecell[lt]{\textbf{SD Search}\\\textbf{Time (s)}} & 
\makecell[lt]{\textbf{Mean}\\\textbf{Fixations}} & 
\makecell[lt]{\textbf{SD}\\\textbf{Fixations}} &
\makecell[lt]{\textbf{MAPE}\\\textbf{(\%)}} \\
\midrule
16 & 1.3 & 0.8 & 7.7  & 4.6  & 23.2 \\
24 & 2.0 & 1.3 & 11.6 & 6.8  & 17.6 \\
48 & 4.0 & 2.6 & 23.6 & 14.0 & 0.6 \\
\bottomrule
\end{tabular}
}
\caption{Search times, number of fixations, and the mean absolute percentage error (MAPE) for the baseline simulation with unlimited memory across all set sizes.}
\label{tab:tab_base}
\end{table}

\begin{table}[H]
\vspace{-2pt}
\centering
\resizebox{\columnwidth}{!}{%
\begin{tabular}{l l l l}
\toprule
\textbf{Semantic order} & \textbf{Set Size} & \textbf{Mean Fixations} & \textbf{SD Fixations} \\
\midrule
No categorization & 16 & 8.7 & 4.7 \\
No categorization & 24 & 12.7 & 6.9 \\
No categorization & 48 & 24.1 & 14.1 \\
Shared            & 16 & 7.2 & 4.1 \\
Shared            & 24 & 10.7 & 6.2 \\
Shared            & 48 & 21.4 & 12.3 \\
Unique            & 16 & 6.5 & 3.7 \\
Unique            & 24 & 9.5 & 5.5 \\
Unique            & 48 & 18.2 & 10.7 \\
\bottomrule
\end{tabular}
}
\caption{Mean and standard deviation of the number of fixation predictions by the adaptive model across semantic order and set size conditions.}  
\label{tab:mod_adap}
\end{table}

\subsection{Discussion}
The experiment replicated the well-established finding that visual search time increases with the number of elements \citep{treisman1980feature, wolfe1994visual}.
Moreover, it replicated the previously reported finding that visually consistent semantic grouping aids in visual search \citep{hornof2001visual, halverson2011computational, salmeron2005expert, chen2015emergence, bailly2014model, ahlstrom2010s, cockburn2007predictive, halverson2008effects, card1982user, mcdonald1983searching, halgren1993towards, brumby2015visual}.
As a finding that to our knowledge has not been extensively investigated and reported on, we show an interaction: the positive impact of good semantic design on search time increases as set size increases.
From a design perspective, this implies that the alignment between semantic and visual grouping becomes more critical as information load increases.
Although our model slightly overestimates search times for larger set sizes, the experimental results still demonstrate its ability to accurately predict the effects of different conditions across a wide range of set sizes.
The adaptive model's fixations are close to the theoretical minimum for encoded elements during the task. 
This supports the adaptation of both the model and humans in utilizing spatial and semantic structures to minimize revisits and efficiently reduce the search area.
This speaks for the power of computational rationality: By implementing a psychologically valid internal environment and an external environment that corresponds to the human task, human-like behavior emerges as an optimal adaptation to these environments.

The first experiment confirmed that our model can account for human-like adaptation to visual and semantic structures in a search task in terms of aggregate search times.
This adaptation is posited to arise from humans' strategic ability to hierarchically inhibit return during visual search, thereby augmenting the limitations of VSTM.
However, while aggregate search times are a useful metric for usability of visual designs, it lacks detailed analysis of eye movement patterns.
To address this, a second experiment focuses on comparing human and model-simulated aggregate eye movement patterns.

\section{Experiment 2}\label{sec5}
\subsection{The Dataset}
\begin{figure}
    \centering
    \includegraphics[width=0.5\textwidth]{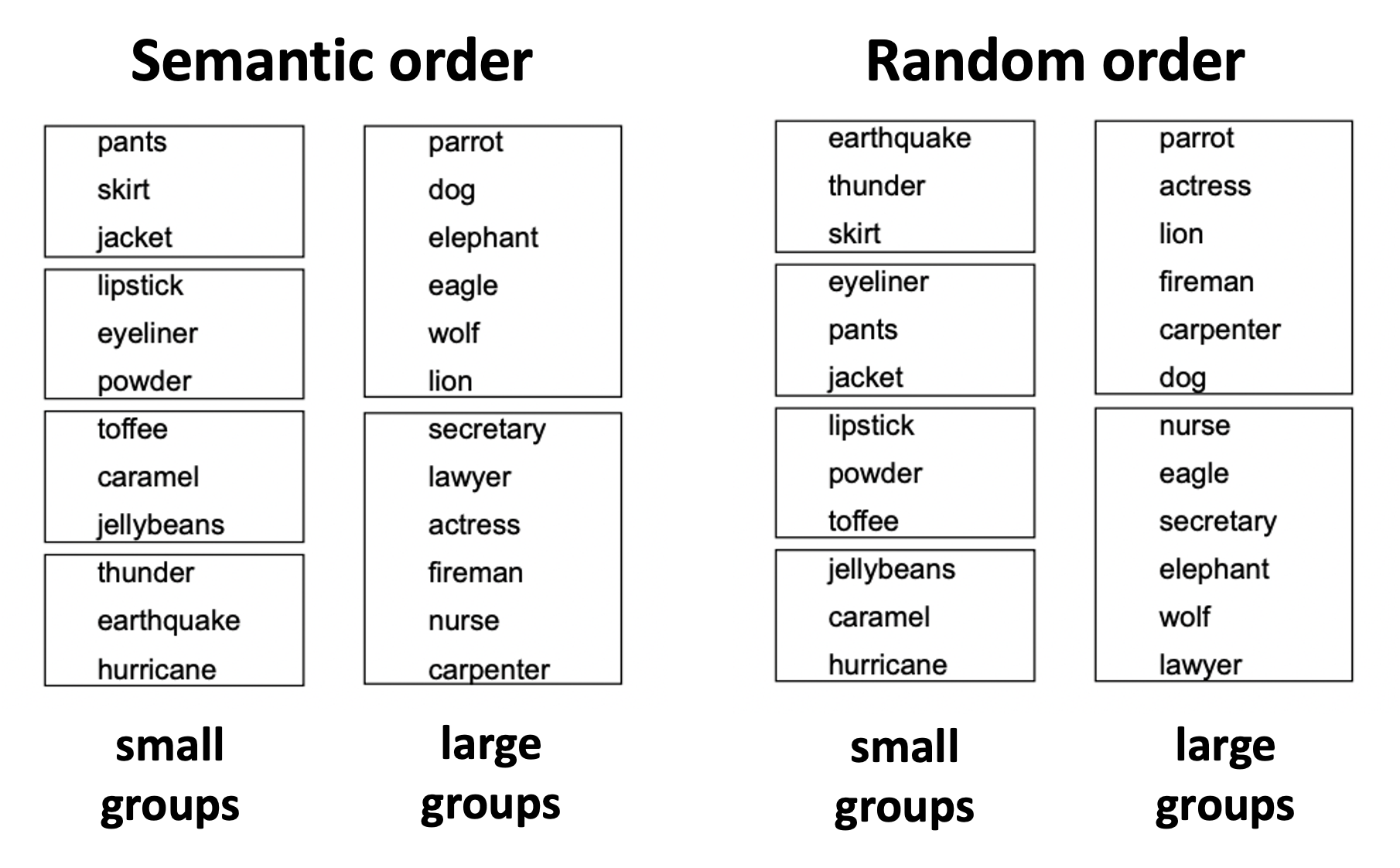}
    \caption{Conditions for the second experiment from the study by \citet{brumby2015visual}}
    \label{fig:fig_conditions}
\end{figure}

The second experiment utilizes data from a previous menu selection study conducted by \citet{brumby2015visual}, where participants engaged in visual search for cued targets across various menu layouts.
The study employed a $2 \times 2 \times 2$ mixed factorial design, which included variables for visual order, group size, and visual grouping.
Visual grouping was a between-subjects variable, while the other factors were within-subjects.
Each layout comprised 36 single-word items categorized into an everyday semantic category—such as animal, building, or entertainment, further divided into four subcategories.
A total of $N=36$ participants completed 100 visual search trials each.
The layouts were organized into structured visual groupings (small or large) and were either semantically or randomly ordered, akin to the conditions in our first experiment (Figure \ref{fig:fig_conditions}).

The original study included a condition that visually delineated groups from each other.
Since our model does not simulate the low-level vision required to perceive such features, we have omitted this condition.
We trained the model on tasks similar to those in Experiment 1 and simulated the results with 5000 visual search tasks per condition (details provided in Appendix B).
To assess the model's accuracy in predicting task performance and fixation count we report $R^2$ and $RMSE$ based on aggregate metrics reported in the original study.
In addition, we compare aggregate ordinal item jumps (saccade distances), and within group visits and number of group visits predicted by our model to the human values.
Notably, the saccade distance in the original data aligns most directly with the metric termed ordinal item jumps in our results.

\subsection{Results}
\begin{figure}[t]
  \centering
  \includegraphics[width=0.5\textwidth]{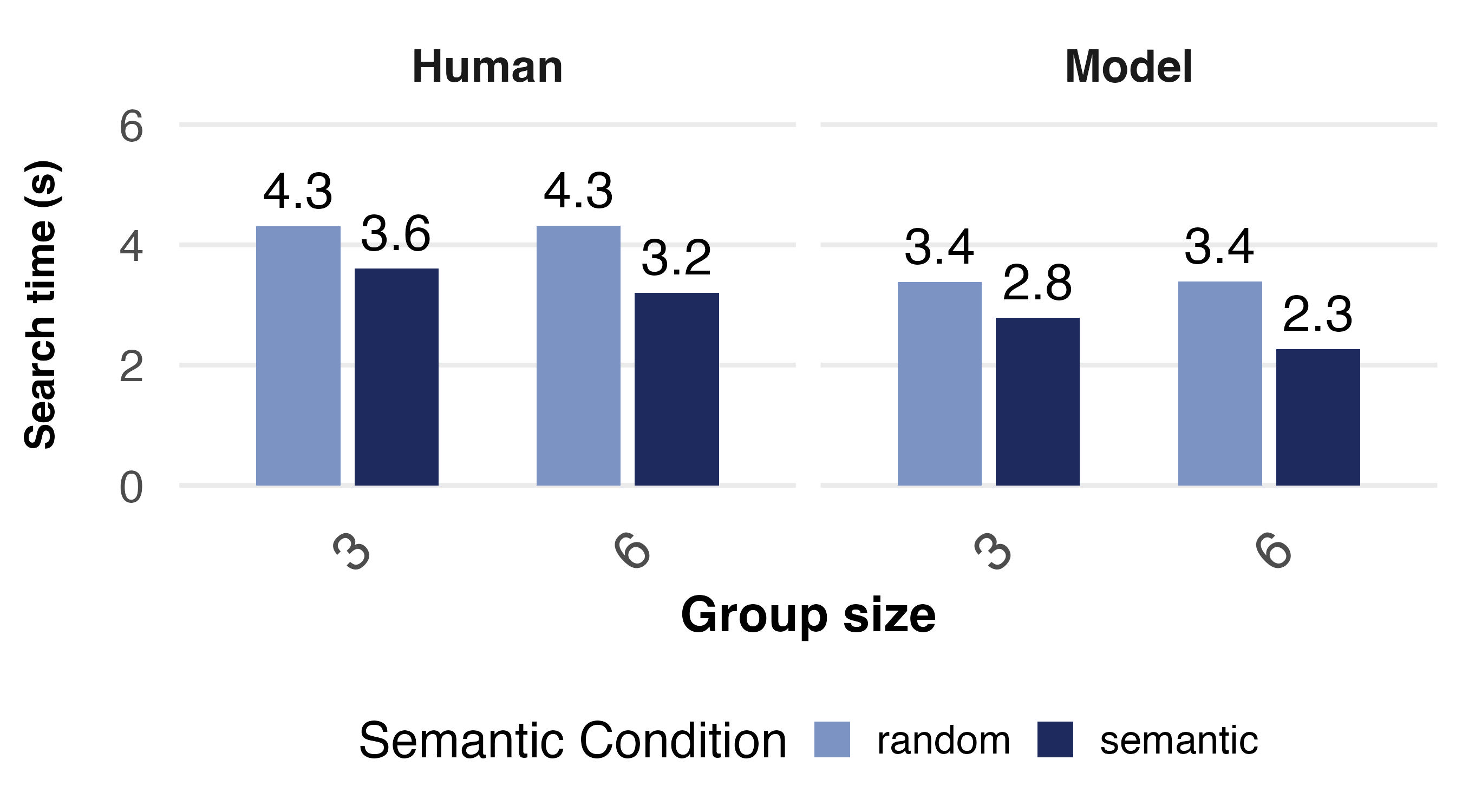}
  \caption{Average Search time between semantic condition and group size, for human and model data.}
  \label{fig:fig_menu_time}
\end{figure}

\begin{figure}[htbp]
    \centering
    \includegraphics[width=0.5\textwidth]{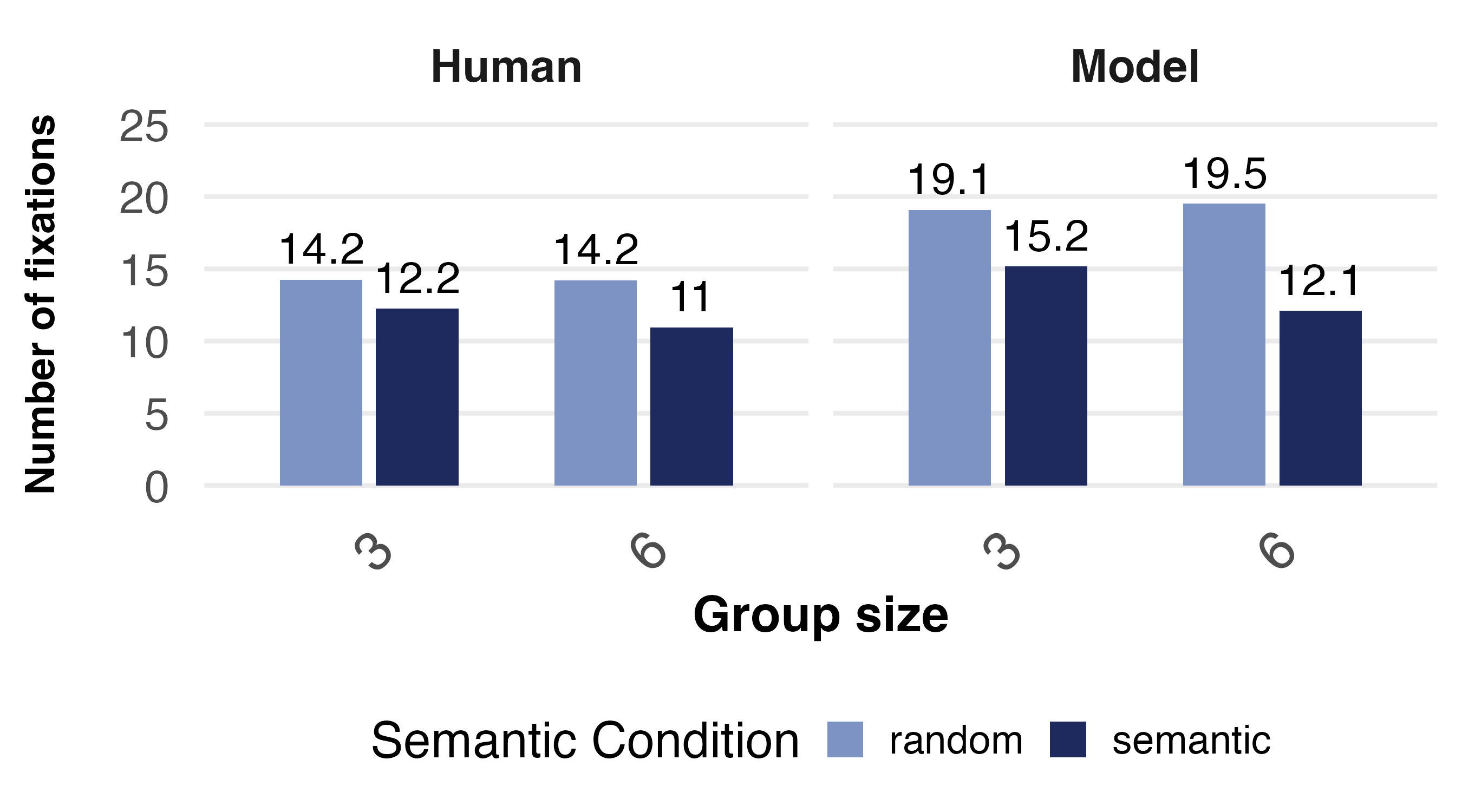}
    \caption{Average fixation count between semantic condition and group size, for human and model data.}
    \label{fig:fig_menu_fix}
\end{figure}

Figure \ref{fig:fig_menu_time} presents the average search times for both humans and the model under various conditions. In addition, Table \ref{tab:menu_base} shows the number of fixations and the percentage error (MAPE) compared to human search times for the adaptive and baseline models.
Semantic order improves search times over random order for both humans and the model.
Group size also has, as expected, a reducing effect on search time. 
Although the model’s search time is lower, the model closely matches human performance, with $R^2 = 0.99$ and $RMSE = 0.05s$.
The regression, based on four data points from systematically controlled conditions, effectively captures meaningful contrasts despite the limited dataset size.
Average fixation counts are presented in Figure \ref{fig:fig_menu_fix}. 
The results reflect trends similar to search times, with fewer fixations in semantically ordered menus.
The model accurately predicts this effect on fixation, with $R^2 = 0.99$ and $RMSE = 0.15s$.
Fixations are systematically higher for the model; however, the results show relatively similar differences between the conditions in both the model and the human data.

\begin{table}
\centering
\resizebox{\columnwidth}{!}{%
\begin{tabular}{l l l l l l}
\toprule
\makecell[lt]{\textbf{Model}} & 
\makecell[lt]{\textbf{Group}\\\textbf{Size}} & 
\makecell[lt]{\textbf{Semantic}\\\textbf{Order}} & 
\makecell[lt]{\textbf{Search}\\\textbf{Time (s)}} & 
\makecell[lt]{\textbf{Number of}\\\textbf{Fixations}} & 
\makecell[lt]{\textbf{MAPE}\\\textbf{(\%)}} \\
\midrule
Adaptive & 3 & Random   & 3.4 & 19.1 & 21.6 \\
Adaptive & 3 & Semantic & 2.8 & 15.2  &  22.8 \\
Adaptive & 6 & Random   & 3.4 & 19.5 &  21.5 \\
Adaptive & 6 & Semantic & 2.3 & 12.1  & 29.1 \\
Baseline & - & Random   & 3.1 & 17.4 &  28.9 \\
\bottomrule
\end{tabular}
}
\caption{Predictions of the Adaptive and Baseline models (search time, number of fixations, MAPE) categorized by group size and semantic condition.}
\label{tab:menu_base}
\end{table}

\begin{figure}[H]
  \centering
  \includegraphics[width=0.5\textwidth]{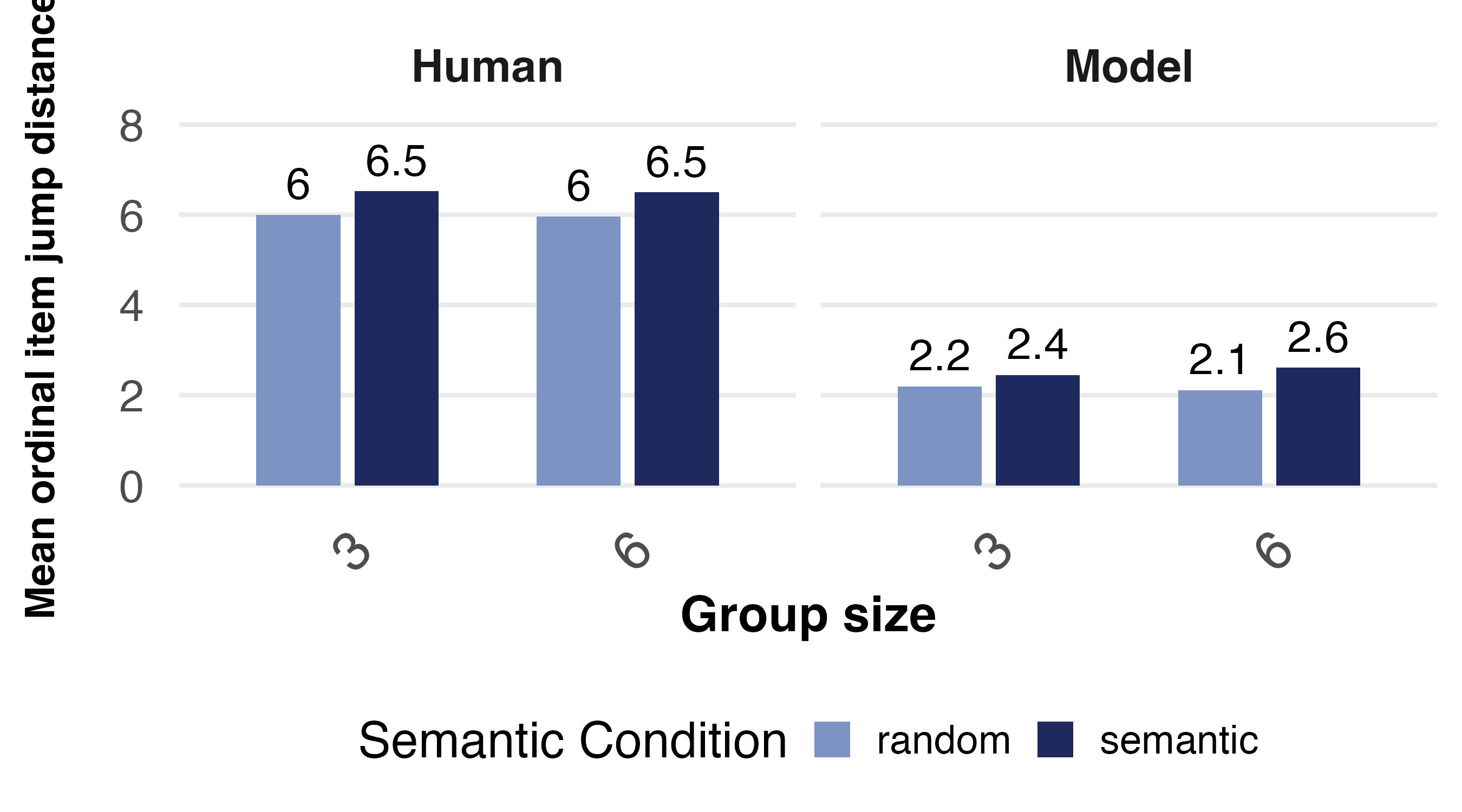}
  \caption{Average ordinal item jumps between semantic condition and group size, for human and model data.}
  \label{fig:fig_menu_saccade}
\end{figure}

Figure \ref{fig:fig_menu_saccade} presents the average number of item skips between fixations, relative to the order of items from the top left downward and then row by row to the right.
In other words, this measure indicates the extent to which the search path deviates from the shortest possible path.
The model fit for this measure is fairly good ($R^2 = 0.89$, $RMSE = 0.09\,\text{items}$). 
With semantic structure, saccades are longer than in the random condition, and the differences are highly consistent also with respect to group size, although in the model the item-to-item saccades are overall considerably shorter.

\begin{figure}[H]
  \centering
  \includegraphics[width=0.5\textwidth]{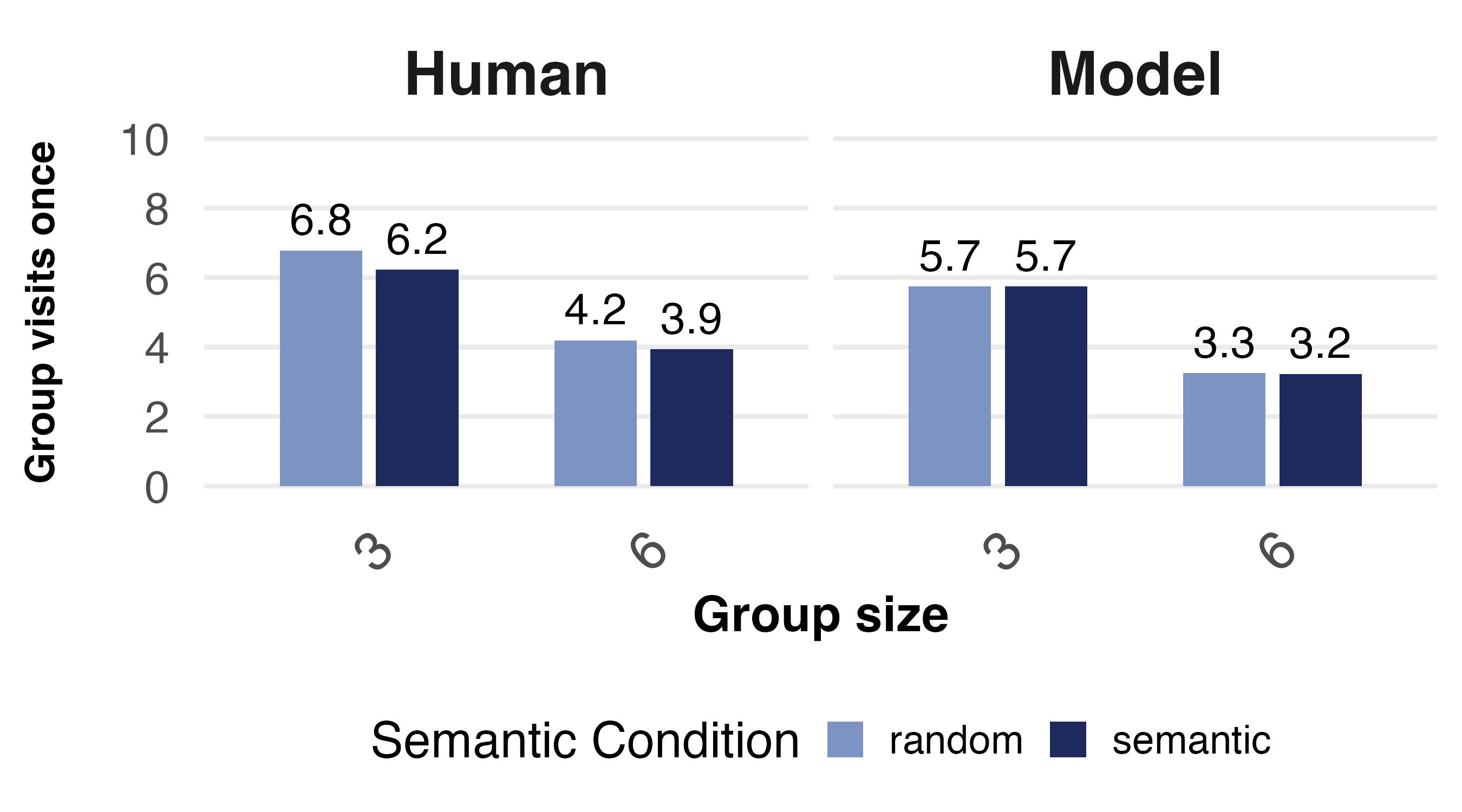}
  \caption{The average number of group visits, when a group has been visited only once during the task, excluding last visited group, for each condition.}
  \label{fig:fig_group_vis}
\end{figure}

\begin{figure}[H]
  \centering
  \includegraphics[width=0.5\textwidth]{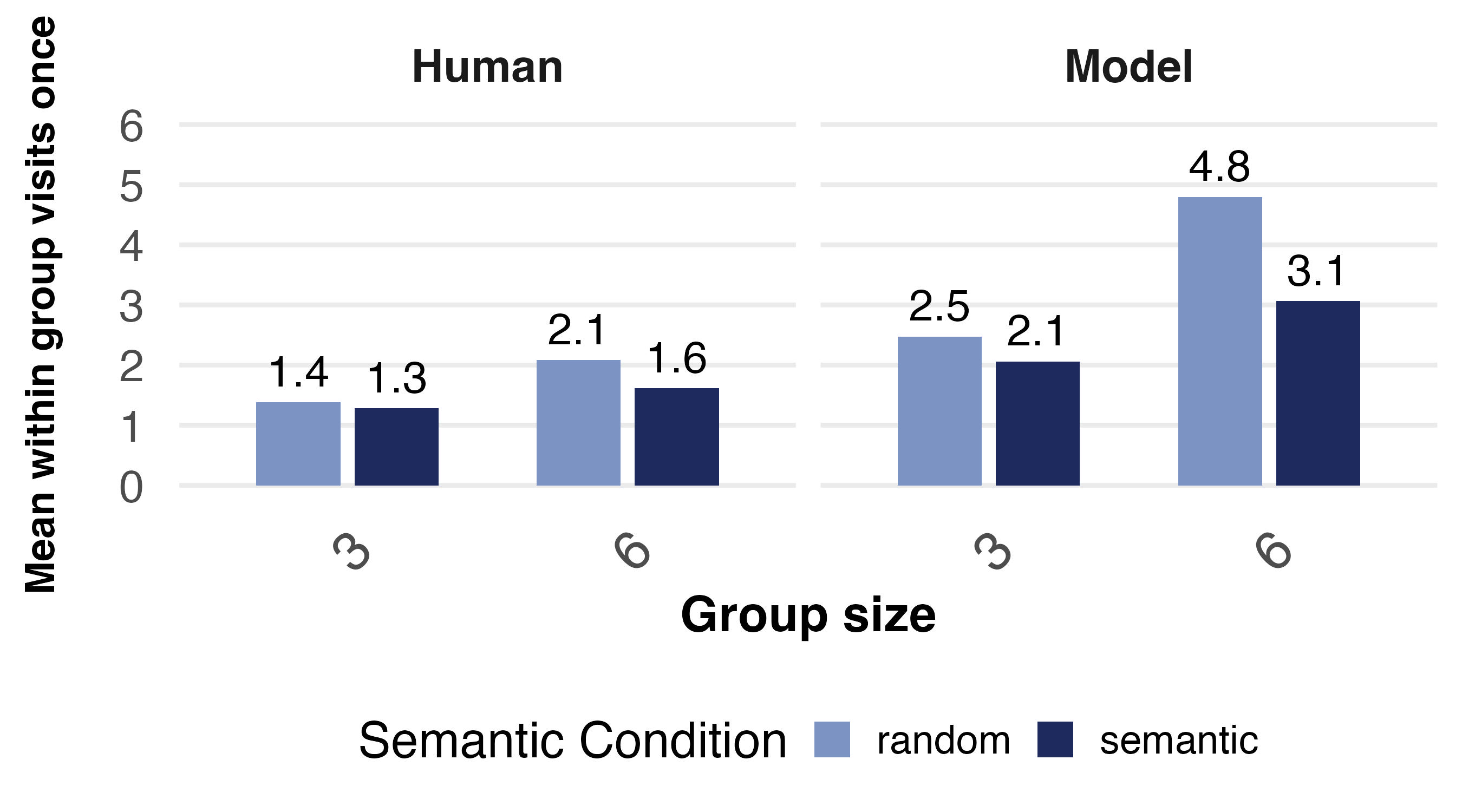}
  \caption{The average number of item visits within groups that were visited only once during the task, excluding target group, for each condition.}
  \label{fig:fig_within_vis}
\end{figure}

The Group visits once metric quantifies the average number of groups that were visited exactly once during the task. 
As shown in Figure \ref{fig:fig_group_vis}, the differences between conditions are of similar strength for both the human data and the model, with $R^2 = 0.97$ and $RMSE = 0.02,\text{visits}$.
In the model, there is no effect of group size, and in the human data the effect is small.

Figure \ref{fig:fig_within_vis} presents the number of item visits within groups that were visited only once during an individual search task, excluding the target group.
This resulted in a close fit to the data, with $R^2 = 0.99$ and $RMSE = 0.02\,\text{visits}$.   
This metric reveals fewer visits in the semantic condition for both group sizes, although the magnitude of the difference varies in the same way as with the other metrics we reported.

In Experiment 2, we also tested a baseline simulation (Table \ref{tab:menu_base}), using unlimited memory and a search strategy where attention always shifts to the nearest next element.
This baseline simulation yielded results close to those of the adaptive model (an average of $3.1$ seconds and $17.4$ fixations), demonstrating the model’s ability to efficiently learn to utilize spatial and semantic information in a human-like manner

\subsection{Discussion}

Our modeling results reinforce the previous finding on the search strategy for spatially and semantically organized items \citep{brumby2015visual} and explain how this search strategy emerges through adaptation.
The inclusion of eye movement data further validates our model.
Semantically organized layouts decrease fixation count, improving search efficiency and target acquisition speed.
Although the search performance of the model is superior to that observed in the eye-tracking data (Figure \ref{fig:fig_menu_time}), it effectively captures the influence of semantic order and group size.
Larger categorized groups facilitate faster search by allowing the model to bypass a greater number of irrelevant items in a single fixation.
In contrast, the random order condition consistently requires exhaustive item-by-item inspection within each group, regardless of group size.

Our model predicts the differences between conditions quite accurately across all presented metrics with fitting only one free parameter: the constant eye movement preparation time in EMMA.
The results strongly support the emergence of a hierarchical search strategy as a result of adaptation to semantic grouping.
In addition to search times, the remaining metrics further support the model’s ability to predict the advantage of semantic categorization in enabling the search strategy to efficiently inhibit irrelevant list groups using semantic information.
First, the number of fixations is lower under semantic organization, indicating a shorter search path.
Secondly, the ordinal item jump distance increases with categorized grouping, since users can skip entire groups with fewer short within-group fixations and longer between-group fixations.
Thirdly, the number of within-group fixations that were visited only once provides further evidence that semantic grouping enables groups to be processed with fewer encodings.
Furthermore, the within-group visits result supports our model’s assumption that humans are unable to skip all irrelevant semantic groups in a single encoding perfectly. 
Exploiting category information is constrained even under the semantic condition, so adaptation relies on an alternative strategy—specifically, a structured search strategy.

Our model replicates the main effects on all metrics, but the magnitude of the results has differences with the human data.
This is because the higher fixation count of the model reflects its aim to simulate attentional encoding rather than to replicate physical eye movements, whereas a human observer is capable of encoding multiple items within a single fixation \citep{cavanagh2005tracking, ojanpaa2002eye}.
This interpretation is further supported by the number of groups visited only once during the task.
Despite differences in other metrics, the magnitude of group visits remains comparable between humans and the model.
Additionally, in the random condition, the baseline simulation still produces a higher number of fixations than the human results, given that the model has perfect memory.
The model’s inherently higher fixation density results in shorter intervals for the ordinal number of item jumps and increases the number of groups that are visited only once within the task.
However, the model can be assumed to slightly overperform relative to humans, as it does not account for all of the factors influencing human attentional shifts.

The results of the second experiment indicate that the hierarchical organization of visual memory enables the emergence of a global search strategy based on semantically structured hierarchies. 
Within the scope of the present experiments, our model accounts for a relatively small item sets and for a restricted number of levels within the visual hierarchy.  
Previously, authors of a model of visual search proposed that future research should indeed explore the influence of hierarchical representations on complex search tasks, such as GUIs with tens of even a hundred elements \citep{jokinen2020adaptive}.
This paper is a step forward to this direction.

%\section{UI simulations}
%\input{06-ui-simulations-3}

\section{General Discussion}\label{sec6}
\textbf{Overview}:
We present a model that simulates visual search adaptation to coherent visual and semantic layout structures, providing improvements over existing models \citep{cockburn2007predictive, bailly2014model, teo2012cogtool, halverson2011computational}.
Central to our model is the principle of computational rationality, which explains eye movement and visual search behavior as optimal adaptations to the constraints of vision and the task environment.
%A key insight of our model is that the human visual system employs hierarchical representations to simplify the search task, thereby mitigating the limitations of visual short-term memory (VSTM).
A key insight of our model is that the human visual system adaptively employs hierarchical representations to simplify search tasks, effectively mitigating the limitations of visual short-term memory (VSTM).

Our model's validity is supported by two studies.
The first, original to this paper, provides detailed data on how semantic grouping and set size interact.
As layout complexity increases, the importance of effective visual and semantic design becomes important.
The second experiment investigates findings from an earlier study \citep{brumby2015visual}, which showed that semantic structures impact eye movements and improve performance.
Our results go beyond this by revealing how search strategies emerge from such an environment.
Our model successfully replicates the impact of semantic grouping on fixation count, item jump distance, and within-group visits as well.

%in our model, the benefit of semantic grouping results from rational decision-making, rather than purely from assumptions about the decision-making logic of the search strategy or relying on mathematical functions derived from empirical data. The predictive accuracy of our model stems from its independent learning process, which accounts for the impact of cognitive constraints and environmental limitations. The versatility of our model is demonstrated by its ability to replicate empirical results across a wide range of item sizes. Additionally, the model enables predictions for any visual grouping combined with semantic grouping, regardless of whether the grouping is presented as a menu layout or a spatial list arrangement. Possibly, visual grouping by other visual features may not produce a difference in the prediction, but our study focuses solely on spatial proximity and grouping by boundaries.

\textbf{Limitations}:
This paper focuses on the optimal adaptation of visual search to hierarchical memory structures within semantically grouped layouts.
It does not implement a model of how visual groups are formed within visual representations or how semantic distances between labels are computed.
At the beginning of our study on understanding adaptability to semantic hierarchies, we deliberately highlight a simplified framework for the semantic relations of UI environments.
However, in practice, these relations are likely to be considerably more complex in human representations.

Future work could utilize computer vision to derive visual structures from raw stimuli, and word embeddings to model semantic knowledge, as long as these models are psychologically validated.
Moreover, our experiments indicate that, particularly in layouts with a large number of elements, our model overestimates search time and fixation count relative to human performance.
This discrepancy suggests that humans utilize additional hierarchical structures not accounted for in our current model.
While a full understanding of the human visual representation remains a goal for future research, this study successfully models how visual search adapts to the presence of such hierarchical representations.
Finally, our model does not identify parameters for simulating how individual differences, such as visual limitations or task motivation impact visual search behavior.
As such parameters have been identified with respect to visual search \citep{jokinen2021touchscreen,sarcar2018ability}, a future extension to our model could address such individual differences.

%The model's dynamic limitations in visual memory may possibly explain why the model's learning lagged behind human search performance at the largest set size (48 items), when the layout was not categorized but visually grouped. Similarly, in Brumby and Zhuang's semantic order condition, model’s search time increased with larger visual grouping. This memory limitation was already stated in our previous model (LÄHDE VANHAAN MALLIIN), where the real bottleneck of the model lies in the discrete memory capacity and the constraints of hierarchical levels. Additionally, the model's visual and semantic grouping is symbolically predefined information, which makes the internal representation partially static in nature. It is far more probable that cognition can dynamically create smaller memory chunks during the serial perceptual process, as well as enlarge existing perceived structures and form new chunk levels from smaller chunks. Thus, memory supports the efficiency of more flexibly saving short-term memory for only the necessary locations where items have already been scanned.

\textbf{Implications}:
To facilitate open science, we publish all data and code.
Our research improves our understanding of visual search over layouts, a topic extensively studied in HCI \citep{hornof2001visual, halverson2011computational, salmeron2005expert, chen2015emergence, bailly2014model, ahlstrom2010s, cockburn2007predictive, halverson2008effects, card1982user, mcdonald1983searching, halgren1993towards, brumby2015visual, jokinen2020adaptive}.
We advance knowledge of how human vision adapts to complex UI wireframe design configurations, addressing a gap previously identified in vision modeling \citep{jokinen2020adaptive}.
Through the adaptability of the model, designers gain a better understanding of how to account for the limitations of visual memory when organizing semantic content alongside visual structures, particularly as item size increases.
%The model introduced here can also serve as a tool for designers to understand the impact of their design decisions on user behavior.
The model introduced here can also serve as a tool for designers to understand the impact of their design decisions on user behavior. 
This plays a crucial role across various types of user interfaces, such as list groupings, tab menus, spatially organized grids, and dropdown menus.
In its current state, the model can process a wireframe UI design and semantic information about element labels to simulate step-by-step eye movements and information processing in the VSTM during visual search.
Designers can use this to learn, compare designs, or even optimize design candidates.
This proves particularly valuable in scenarios where straightforward semantic structures would be impractical, assisting designers in balancing semantic design with other objectives.

%The primary potential application of the model would be as a tool for UI designers to evaluate user search performance for so-called wireframe design. Ideally, this would mean that once the designer has defined a very simple structural UI design with the locations of items and their respective semantic and visual groupings. The model would then cover also visual representations that convey semantic content, such as icons. In order to develop the model into a practical application, visual grouping algorithms could be integrated into the model to simulate low-level visual processes for perceiving visual structures, as well as evaluating semantic distances for semantic grouping, using token embeddings data. Additionally, the model could potentially be evolved into a tool for design solutions that optimizes the visual hierarchy of structures to align with semantic content, guided by reinforcement learning algorithms.

\textbf{Conclusion}:
In recent years, cognitive models based on computational rationality have emerged as the new frontier in modeling interaction \citep{chandramouli2024workflow, chen2015emergence, jokinen2020adaptive, fu2007snif, jokinen2021multitasking, jokinen2021touchscreen, oulasvirta2022computational, todi2019individualising, wang2025pedestrian}.
Our model aligns with this approach, demonstrating that adopting the strong prior assumption of human cognition optimizing behavior policies within the constraints of itself and the task environment leads to models that simulate human-like behavior using a minimal number of free parameters.
Model validation involved adjusting only a constant time parameter and the EMMA preparation parameter in each tasks, to account for interactions outside the scope of our visual search model.
All other parameters were either based on earlier psychological literature or objective task designs.
Looking ahead, we envision architectural solutions that integrate lower-level models like ours with higher-level models of complex interaction, illustrating how different aspects of human cognition coadapt for efficient interaction.

%% Preliminary results highlight the value of semantic organization combined with visual grouping.  However, from the perspective of HCI research, the question remains broader for the future:
%% How does the alignment of visual and semantic hierarchies affect cognitive constraints in employing information? This is crucial from a UI design perspective, as design fundamentally begins with the question of how semantic information should be visually structured. However, it is still unclear what the limits of this alignment are, specifically when visual relationships create consistency with hierarchical information and when they undermine it.

%% Our model's rational decision-making approach significantly enhances the understanding of visual search, particularly in how decision-making learns to utilize semantic information in visually structured environments. It provides a deeper understanding of the emergence of various search strategies and supports behaviorally consistent assumptions regarding the use of cognitive resources in constrained interaction environments. This also paves the way for increasingly accurate simulations of user behavior in the future, as we gain a deeper understanding of cognitively limiting factors in visual task and how to model them computationally.

\section{Data availability}
An anonymized repository \citep{anonymous2025osf} provides the raw and processed datasets for Experiments 1 and 2 as well as all analysis scripts. The model implementation/code is also hosted in the same repository.

%%%%%%%%%%%%%%

\begin{appendices}
\section{EMMA model}\label{sec7}
We use the EMMA model to simulate eye movements \citep{salvucci2001integrated}. Time to encode a visual object is
$ T_e = K \cdot \left[ -\log(f) \right] \cdot e^{k \cdot \epsilon}$,
where $f$ is the frequency of the encoded element set to 0.1 here;
$K$ (0.006) and $k$ (0.4) are scaling constants; and
$\epsilon$ indicates eccentricity or the distance between the current fixation location and the target element in visual degrees.
Given eccentricity EMMA predicts whether a saccadic movement is necessary to refixate.
In the tasks modeled here, element distances are large enough to always require this.
Refixation duration is $T_s = t_{\text{prep}} + t_{\text{exec}} + D \cdot t_{\text{sacc}}$,
where $T_{\text{prep}}$ (0.135) is constant time to prepare the saccade;
$T_{\text{exec}}$ (0.07) is constant time to execute it;
$t_{\text{sacc}}$ (0.002) is scaling constant for saccade distance $D$.
All parameters are set to defaults from the original source, where they were fitted to human data.

\section{Agent Representation and Training}\label{sec8}
The model uses proximal policy optimization (PPO), a deep RL algorithm \citep{schulman2017proximal}.
The action space for the PPO agent was discrete($N$), where $N$ is the number of visual elements.
Each action corresponds to fixating on a particular visual element.
The observation space is listed in Table \ref{tab:obs_space}, and it one-hot or multi-hot encodes features of visual elements, with each visual element having a constant index in the one-hot vector.
\texttt{obs\_h} encodes each element in the layout on the basis of whether it belongs into the visual group which is currently being fixated upon.
\texttt{vstm} encodes previously encoded elements, limited to 6, in a first-in-first-out fashion.
\texttt{fixation} encodes the element currently under fixation.
\texttt{sem\_match} encodes all elements that match the target category: if the element under fixation matches the target category and the task condition guarantees semantic grouping, then all elements of this category are encoded as a semantic match, with added noise occurring with probability 20\%. 
Encoded groups remain in memory until the end of the task.
\texttt{sem\_mismatch} one-hot encodes non-matching elements, similarly as with \texttt{sem\_match} but for semantic mismatch.
\texttt{distance} is the distance of all visual elements to the current fixation location (this is the only non one or multi-hot vector).
\texttt{group} assigns each visual element to a particular visual group.

\begin{table}[H]
\centering
\caption{Observation space of the PPO agent. \texttt{N} = number of elements, \texttt{G} = number of groups.}
\label{tab:obs_space}
\begin{tabular}{|l|c|c|c|}
\hline
\textbf{Name} & \textbf{Range} & \textbf{Shape} & \textbf{Type} \\
\hline
obs\_h & \{0, 1\} & (\texttt{N},) & int \\
\hline
vstm & \{0, 1\} & (\texttt{N},) & int \\
\hline
fixation & \{0, 1\} & (\texttt{N},) & int \\
\hline
sem\_match & \{0, 1\} & (\texttt{N},) & int \\
\hline
sem\_mismatch & \{0, 1\} & (\texttt{N},) & int \\
\hline
distance & [0, 1] & (\texttt{N},) & float32 \\
\hline
group & \{0, 1\} & (\texttt{N}, \texttt{G}) & int \\
\hline
\end{tabular}
\end{table}

We used Proximal Policy Optimization \citep{schulman2017proximal}, reinforcement learning algorithm, implemented with the CleanRL library. 
All default parameters were used except for \texttt{learning\_rate}, which was set between 0.001 and 0.00005. 
The model was trained separately for each experiment.
During training, the model performed between 0.5 to 1 million search tasks and training lasted about five hours. The duration of a single episode varied between 30-40 milliseconds on a computing cluster.

\end{appendices}

%\section{Competing interests}

%\section{Competing interests}
%No competing interest is declared.

%\section{Author contributions statement}

%The first author wrote the paper and implemented the computational model and conducted the experiment. The second author acted as the supervisor of the study, contributed to the writing, and reviewed the manuscript.

%\section{Acknowledgments}
%The authors thank the anonymous reviewers for their valuable suggestions. This work is supported in part by funds from the National Science Foundation (NSF: \# 1636933 and \# 1920920).

%\bibliographystyle{plain}
%\bibliography{reference}

%\begin{thebibliography}{10}

%\end{thebibliography}

%USE THE BELOW OPTIONS IN CASE YOU NEED AUTHOR YEAR FORMAT.
\bibliographystyle{abbrvnat}
\bibliography{references}

%% sample for biography with author's image
%\begin{biography}{{\color{black!20}\rule{77pt}{77pt}}}{\author{Author Name.} This is sample author biography text. The values provided in the optional argument are meant for sample purposes. There is no need to include the width and height of an image in the optional argument for live articles. This is sample author biography text this is sample author biography text this is sample author biography text this is sample author biography text this is sample author biography text this is sample author biography text this is sample author biography text this is sample author biography text.}
%\end{biography}

%% sample for biography without author's image
%\begin{biography}{}{\author{Author Name.} This is sample author biography text this is sample author biography text this is sample author biography text this is sample author biography text this is sample author biography text this is sample author biography text this is sample author biography text this is sample author biography text.}
%\end{biography}

\end{document}